\documentclass[doublecol]{epl2}
\usepackage{subfigure}
\usepackage{graphicx}% Include figure files
\usepackage{dcolumn}% Align table columns on decimal point
\usepackage{bm}% bold math
\usepackage{multirow}
\usepackage{caption}
\usepackage{amsmath}
\revision

\title{Link power coordination for energy conservation in complex communication networks}

\author{Guo-Qiang Zhang} \shortauthor{Guo-Qiang Zhang}
\institute{Institute of Computing Technology, Chinese Academy of
Sciences, Beijing, 100190, P. R. China}


\abstract{ Communication networks consume huge, and rapidly growing,
amount of energy. However, a lot of the energy consumption is wasted
due to the lack of global link power coordination in these complex
systems. This paper proposes several link power coordination schemes
to achieve energy-efficient routing by progressively putting some
links into energy saving mode and hence aggregating traffic during
periods of low traffic load.  We show that the achievable energy
savings not only depend on the link power coordination schemes, but
also on the network topologies. In the random network, there is no
scheme that can significantly outperform others.  In the scale-free
network, when the largest betweenness first (LBF) scheme is used,
phase transition of the networks' transmission capacities during the
traffic cooling down phase is observed. Motivated by this, a hybrid
link power coordination scheme is proposed to significantly reduce
the energy consumption in the scale-free network. In a real Internet
Service Provider (ISP)'s router-level Internet topology, however,
the smallest betweenness first (SBF) scheme significantly
outperforms other schemes.}

\begin{document}

\maketitle
%\begin{keywords}
%energy-efficient Internet, traffic flow, betweenness, link power
%management
%\end{keywords}
\section{Introduction}
Electricity usage of the data communication networks has contributed
to a large, and growing, portion of the overall energy consumption,
becoming a major concern for network
operators~\cite{cutting-bill}\cite{green-internet}\cite{next-frontier}.
However, the lack of global power coordination makes it hard to
achieve highly efficient use of the energy.
%The opportunities for energy
%conservation lie both in the edge and core of the network. In the
%edge of the network, Ethernet with adaptive link rate~\cite{alr},
%smart network adapters~\cite{skilled-in-art,somniloquy}, and network
%connectivity proxies~\cite{next-frontier,network-proxy,proxying} are
%currently introduced to reduce the energy use of peripheral
%networked devices.
%By proxying some routine network activities, these approaches allow the edge devices to go to sleep when they are idle.
The network infrastructure often maintains rich connectivity to
ensure certain degree of redundancy and robustness to cope with peak
load or to accommodate future growth, which is, however, unnecessary
under the circumstance of low traffic load. Since a link's energy
consumption is mainly dominated by its operation states, e.g.,
active or inactive,
%of a router is dominated by the Chassis and the number of Line cards
%that are powered on~\cite{power-awareness},
rather than the carried traffic load~\cite{power-awareness}, it is
very inefficient to keep all the links powered on when traffic load
is low.

%such that when the traffic load drops, more links could be put into
%low energy usage mode, while when the traffic load rises, links
%should be woken up accordingly.

A principle for energy-efficient routing is that the number of links
kept active should be correlated with the traffic load, an
incarnation of the \emph{energy proportionality}
concept~\cite{energy-proportion} in the context of routing.
Measurements show that the Internet backbone traffic exhibits daily
periodicity~\cite{backbone-traffic}. It starts to rise from around
9:00AM and reaches a high level around 2:00PM, then it plateaus
until 2:00AM, after which it starts to decline. Roughly, the
backbone traffic stays at the high level for half of the day, and
the warmup and cooling down phases account for the other half day.
During the long period of traffic decline (or rise), links can be
progressively put into sleep (or woken up) for energy conservation.
Motivated by this, the pioneer work \cite{green-internet} proposed a
sleeping scheme that relies on local traffic profile to achieve
small-scale coordination.

However, presently there lacks a preliminary knowledge of how the
number of active links correlates with the traffic load. This issue
becomes more complicated since researches in the last decade have
revealed that the data communication networks have heterogenous and
complex structures, rather than presumably random topologies
\cite{power-law-99, Barabasi99, evolution-of-internet-and-core,
symbiotic-effect, newman-review,community-structure,
finding-community, triangular-clustering,
overlapping-community,multiscale-community}. It is well known that
routing on complex networks shows quite different characteristics
from routing on random networks \cite{centrality-and-network-flow,
Arenas01, load-distribution, Danila06, Yan06, Zhang07, zhang2010,
opt-net-topology, onset-traffic-congestion}. Consequently, different
link power coordination schemes may have quite different performance
on different networks. Hence, for the purpose of maximizing the
energy savings under periodic load variation,  it is critical to
develop efficient link power coordination schemes for various
networks and reveal insight into how energy consumption relates to
traffic load. The authors of ~\cite{power-awareness} proposed a
centralized coordination algorithm by formulating it as a
mixed-integer optimization problem, however, the computation
complexity prevents it to be scaled to large networks. In
\cite{nsdi-2008}, the authors proposed rate adaptation and sleeping
strategies based on local monitoring of the traffic to save
energies. In \cite{energy-aware-traffic-engineering}, a distributed
energy-aware traffic engineering scheme is proposed to maximize the
energy savings. This work bases its coordination decision by solving
an optimization problem constrained by the traffic demands and link
utilizations, and proposes practical protocol-level detail of the
traffic engineering. Similar optimization-driven approaches with
traffic matrices as constraints are also proposed in
\cite{power-down, speed-scaling, icc-2009, greencomm-2009}. In
\cite{infocom-10}, a green OSPF protocol is proposed to save energy
during the low traffic period by shutting down some unnecessary
links. All these schemes, however, are designed for generic
networks. They are not tailored for different networks for
performance improvement. In comparison, our work leverages the
long-term traffic periodicity observed by historical traffic
profile, and focuses on an in-depth investigation of how different
network topologies inherently affect possible energy savings. As a
result, our work reveals insight into future network designing. In
addition, our link powerdown schemes are centralized algorithms that
rely on the historical traffic profile and network topologies, which
are fast enough to be applied to large-scale networks.

%In \cite{energy-aware-traffic-engineering}, a distributed
%energy-aware traffic engineering scheme that can leverage both the
%sleeping and rate adaptation abilities offered by future networking
%hardware is proposed to maximize the energy savings without
%sacrificing the performance. This work bases its decision on the
%real-time traffic demand, and proposes protocol-level detail of the
%traffic engineering. In comparison, our work leverages the long-term
%traffic periodicity observed by historical traffic profile, and
%presents a detailed analysis on how different network topologies
%inherently affect possible energy savings. In addition, our link
%powerdown schemes are centralized algorithms that rely on the
%historical traffic profile and network topologies, which are fast
%enough. Their work focuses on energy-aware traffic engineering on
%existing networks, quite useful to be implemented in existing
%networks, whereas our work investigates the relation between network
%topologies and energy-awareness, useful for future network
%designing.

In short, we propose several coordinated link power management
schemes and analyze the efficiency and efficacy in terms of energy
savings on several different network topologies. These schemes rely
on macroscopic topological properties to enable coordination in
large-scale networks. We show that the efficiency and efficacy of
different schemes depend not only on the schemes themselves, but
also on the network topologies. In the random network, except for
the largest betweenness first (LBF) scheme, all other schemes have
similar performance; in the scale-free network, a hybrid scheme
performs much better than other schemes due to the presence of a
phase transition of the network's transmission capacity; and in a
real ISP's router-level topology, the smallest betweenness first
(SBF) scheme significantly outperforms other schemes. In addition to
the understanding of the practical network designing in terms of
possible energy savings, this study can also provide useful inputs
for future energy-aware communication network planning and routing
protocol development.

\section{Traffic Flow Model}
In this paper, we adopt a traffic flow model that is similar to the
model used in \cite{Arenas01,Yan06,zhang2010,
Danila06,Zhang07,load-distribution, opt-net-topology,
onset-traffic-congestion, structural-bottlenecks}. The network is
modeled as a graph $G(V, E)$, where $V$ and $E$ are the set of nodes
and edges respectively, with the number of nodes $|V|=N$ and the
number of edges $|E|=M$. Each node is capable of generating,
forwarding and receiving packets. The shortest path routing is used
for path selection. When there are multiple shortest paths between a
node pair, each time a random selection is made. Each interface is
capable of forwarding one packet at a time step to the next hop.
%edge is assigned with a bandwidth, which is set to forwarding one
%packet at a time step in one direction.
At each time step, $R$
packets are generated with random sources and destinations. When $R$
increases from zero, it is expected to observe a phase transition
from the free-flow phase in which the average number of packets
created and consumed are equal, to the congested phase in which the
number of packets created exceeds what the network can process in
time. The critical packet generating rate is denoted as $R_c$, which
can be quantitatively estimated by the following
equation~\cite{Danila06,opt-net-topology,zhang2010,onset-traffic-congestion,
Yan06, Zhang07, structural-bottlenecks}:

\begin{equation}
R_c=\frac{2N(N-1)}{B_{max}}
\end{equation}
where $B_{max}$ is the maximal edge betweenness
\cite{freeman-betweenness} of all edges. Here, the edge betweenness
$B(e)$ of an edge $e$ is defined as $B(e)=\Sigma_{i\neq
j}\frac{\lambda^{(e)}(i, j)}{\lambda(i, j)}$, where $\lambda(i, j)$
is the number of shortest paths between node $i$ and $j$, and
$\lambda^{(e)}(i, j)$ is the number of shortest paths between node
$i$ and $j$ passing through edge $e$.

In this model, we make a coarse-grained QoS evaluation based on the
overall network's packet injecting rate $R$, rather than a per
source and destination QoS demand. If $R\leq R_c$, we say the QoS is
satisfied, whereas if $R>R_c$, we say the QoS is violated. There are
various other ways to evaluate the QoS, for instance, using a
threshold value $\alpha$ that specifies the maximum utilization of
any link in the network, i.e., the QoS is satisfied only when all
link utilizations are below $\alpha$. In our traffic flow model,
these two evaluation approaches are tightly correlated. $R=R_c$
means some links are saturated, i.e., $\alpha=1$. In order to let
$\alpha<1$, we can scale $R$ by a factor of $\alpha$.

\begin{figure*}[htb]
\centering \subfigure[ER]{ \includegraphics[width=5.8cm]{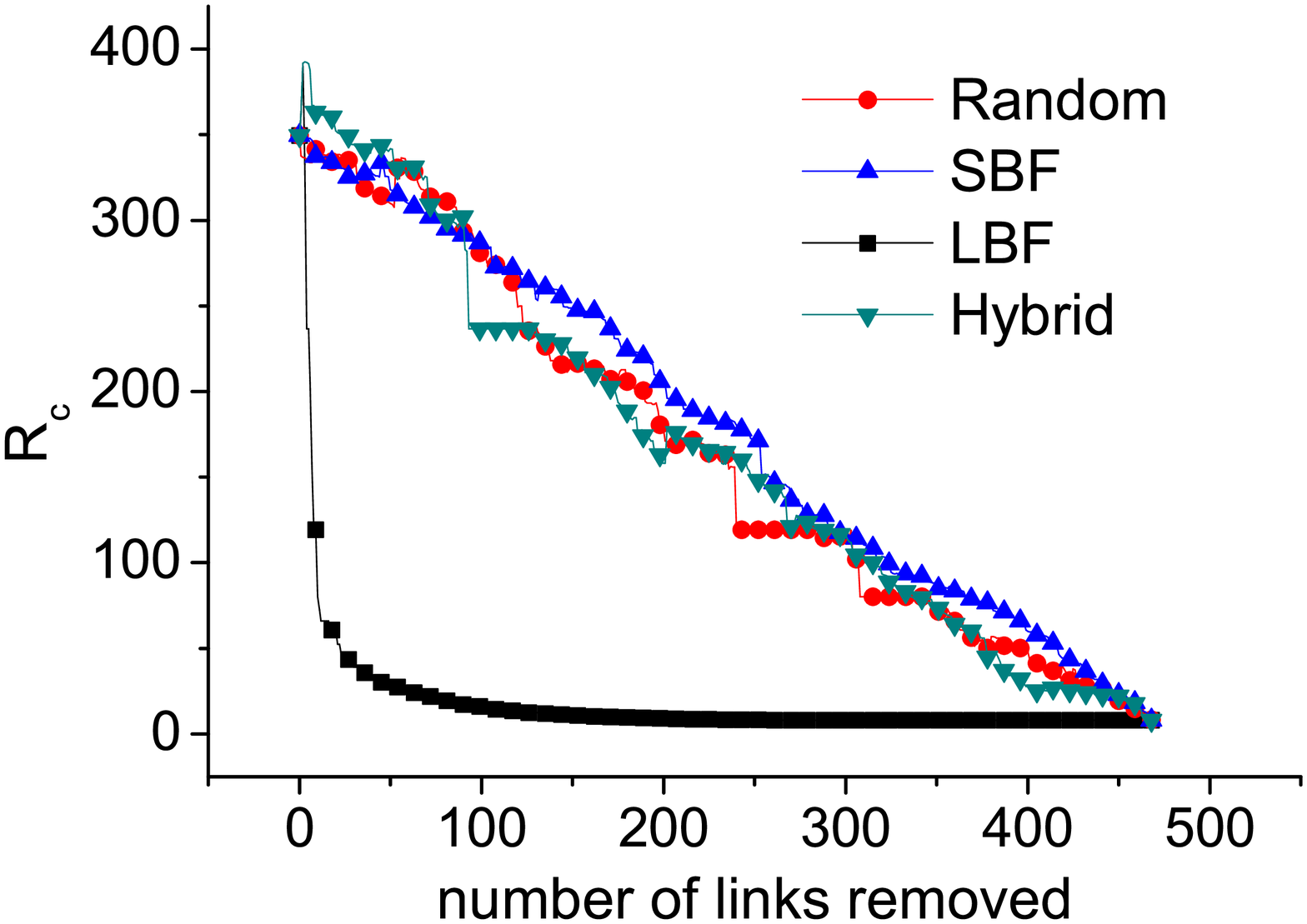}}
\subfigure[BA]{\includegraphics[width=5.8cm]{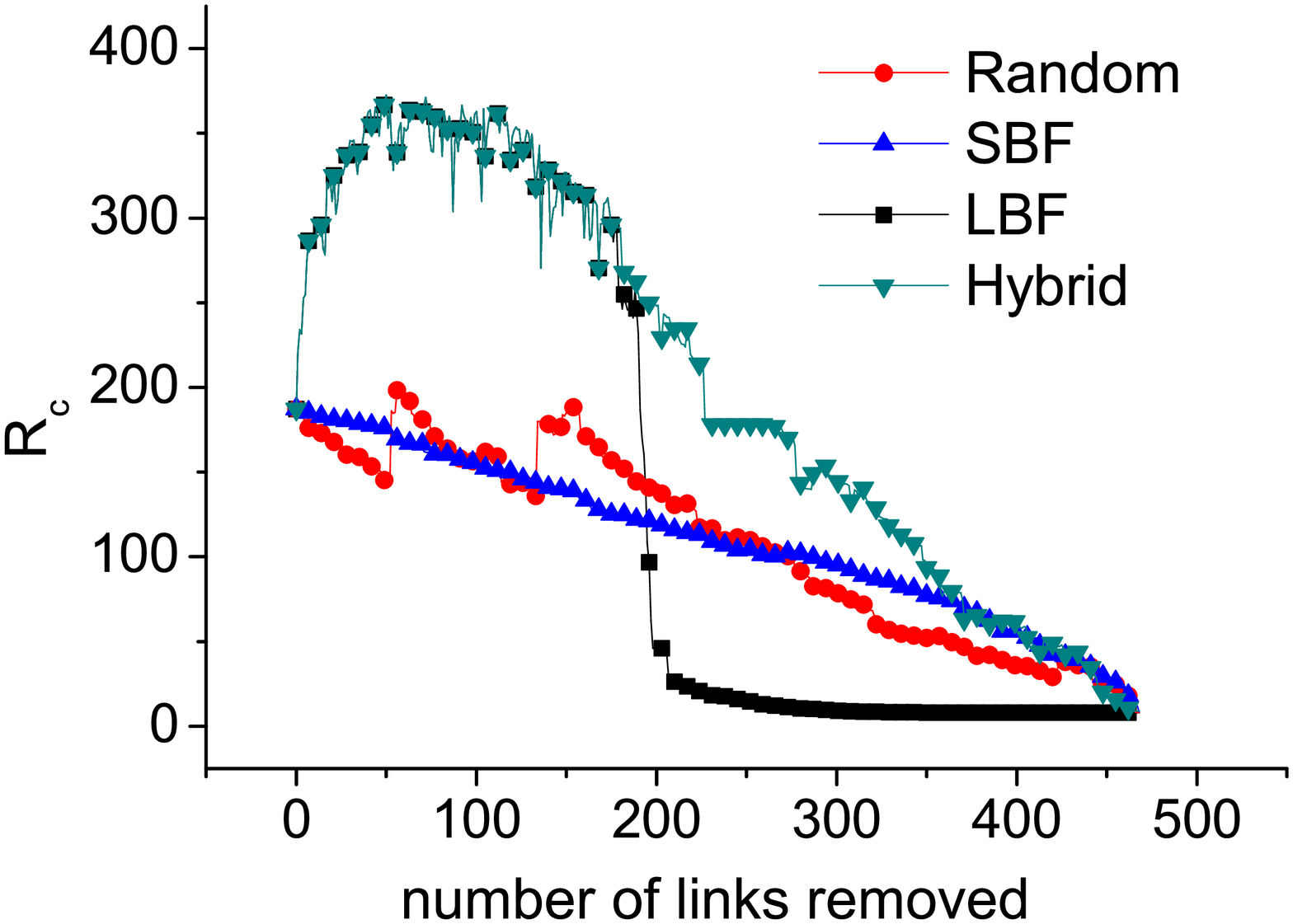}}
%\subfigure[PFP]{\includegraphics[width=8cm]{1c}}
\subfigure[AS3967]{\includegraphics[width=5.8cm]{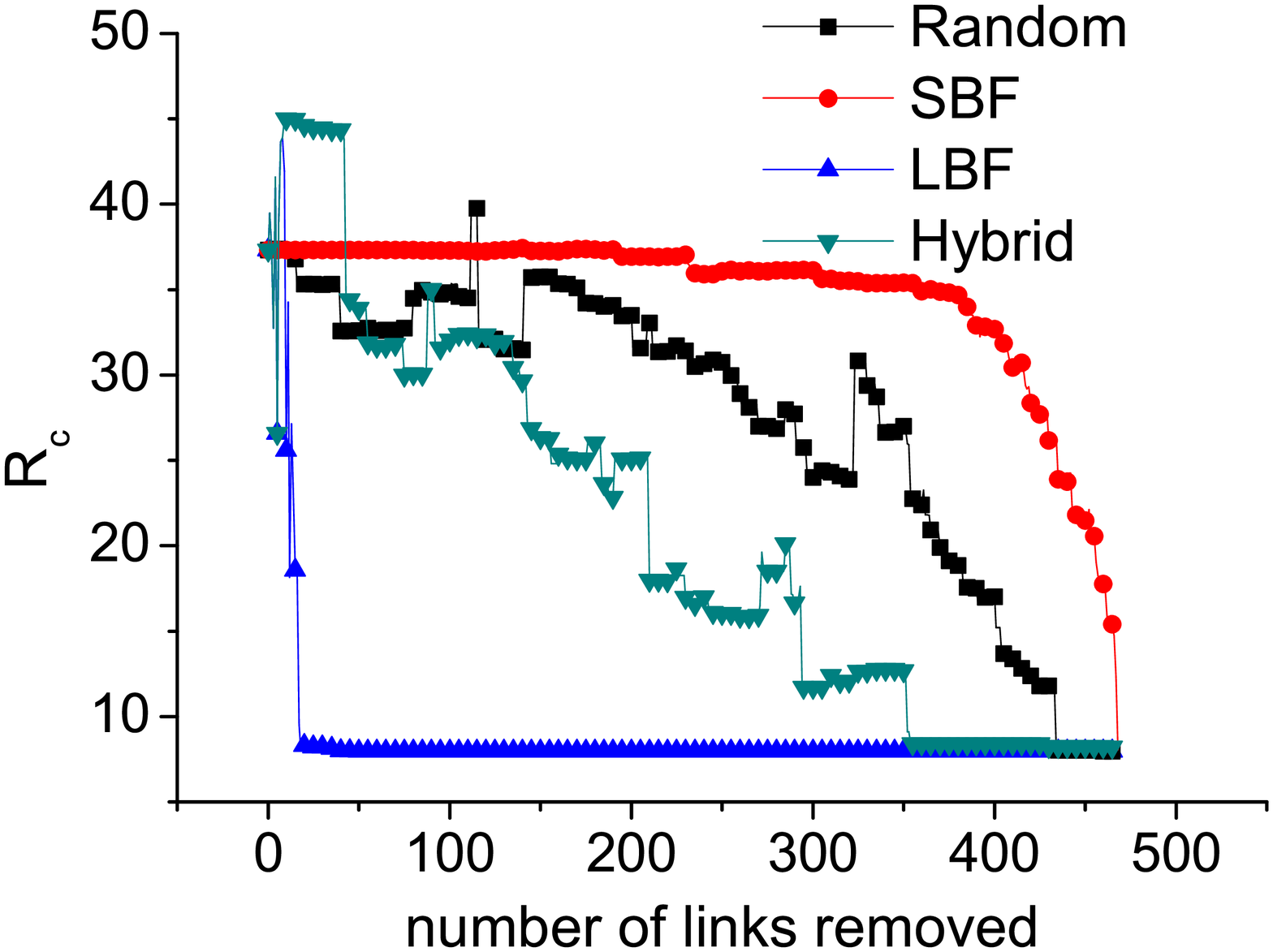}}
\caption{\label{rc}The $R_c$ values after removing network links of
the three networks under different link powerdown schemes.}
\end{figure*}

\section{Link Power Coordination Schemes}

Starting from an original network $G$, we denote $R_c$ of this
original network as $R_0$, and call it as the \emph{designed network
capacity}. The designed network capacity will be used as a baseline
for future quantification of traffic load input, i.e., traffic load
input will be measured by the percentage of the designed network
capacity. A coordinated link powerdown scheme powers down a sequence
of links $\{l_1, l_2, ..., l_h\}$ towards a spanning
tree.\footnote{In order for any node to be able to communicate with
any other nodes, at least a tree containing M-1 links should be kept
active.} We denote the corresponding $R_c$ values of each network
after removing $l_i$ as $R_i$.  We assume a symmetric approach for
link powering up, i.e., when traffic demand rises, links are awoken
in the reverse order as they are powered down. Therefore, in the
following, we will only focus on the traffic cooling down phase.
%\begin{figure*}
%\centering
%\subfigure[ER]{
%\includegraphics[width=8cm]{ER-sleeping}}
%\subfigure[BA]{\includegraphics[width=8cm]{BA-sleeping}} \centering
%\caption{The number of links that can be put into sleep when $R$
%steadily decreases from  $\lfloor R_0\rfloor$ to the minimum of
%$\lfloor R_h\rfloor$. Each time unit, $R$ decreases by one. The
%X-axis represents the elapsed time units since $R$ starts to
%decrease. \label{sleeping}}
%\end{figure*}

We introduce several coordinated link powerdown schemes during the
traffic cooling down phase. The smallest betweenness first (SBF)
scheme comes from the intuition that these links carry the least
amount of traffic, and hence powering down them will not
significantly increase the traffic burden on other links. The
largest betweenness first (LBF) scheme arises from the observation
that the scale-free network's transmission capacity can be
significantly enhanced by the removal of a small number of edges
that connect nodes of high betweenness values \cite{Zhang07}. The
random scheme is included for baseline comparison purpose. The
detailed descriptions of these schemes are listed as follows:

 \begin{itemize}
\item \textbf{Random}: this is the simplest scheme which randomly selects a link and puts it into
 sleep.

 \item \textbf{SBF}: this scheme recursively
powers down the link that has the smallest edge betweenness in the
resulting network. %The betweenness is recalculated each time.

\item \textbf{LBF}: this scheme recursively powers
down the link that has the largest edge betweenness in the resulting
network.
%The betweenness is recalculated each time.

%\item \textbf{Largest betweenness first without betweenness recalculation (LBF-NOREP)}: this
%scheme powers down the link with the largest edge betweenness each
%time, without violating the network connectivity condition. The
%betweenness is calculated only once and used in the subsequent
%rounds.

%In the LBF phase, the betweenness is recalculated each time.
\end{itemize}
In all these cases, we ensure that the outcome network after each
step remains connected. If powering down a link makes the network
unconnected, we will cancel this operation and move to the next
link.

 Fig. \ref{rc}
reports the $R_c$ values after links are removed under different
link powering down schemes of two synthesized network topologies and
a real ISP's router-level topology. The two synthesized networks
are: the random network generated by the ER model
~\cite{random-graphs}\cite{erdos59}, and the scale-free network
generated by the BA model~\cite{Barabasi99}.
%All these synthesized networks are with 600 nodes and an average degree of 6.
The real ISP's router-level Internet topology, denoted as AS3967, is
from AS 3967 measured by the Rocketfule project ~\cite{rocketfuel}.
The synthesized networks are generated with the same size as AS3967,
each having 353 nodes and 820 edges. We run several instances for
each kind of the synthesized networks, and the results are similar.
The results presented here are typical instances of those results.
It is observed that the LBF scheme can increase the $R_c$
significantly when a small fraction of the links are removed in the
BA network, however, there is a phase transition point around which
removing additional links can drastically decrease the $R_c$. With
respect to energy saving, this means a certain number of links can
be put into sleep when the packet generating rate $R$ is smaller
than the $R_c$ before the phase transition, however, after this
point, no links can be put into sleep,  unless $R$ drops below the
$R_c$ of the network after the phase transition, which could be a
quite long time period. Inspired by this, we propose a fourth scheme
called Hybrid, in hope that the $R_c$ can go down gradually rather
than abruptly even after the phase transition, hence creating more
chance for sleeping:
\begin{itemize}
\item \textbf{Hybrid}:
this scheme first uses the LBF scheme to power down links until the
phase transition of $R_c$, after which, it switches to the Random
scheme.
\end{itemize}

The key question in the Hybrid scheme is when to switch from the LBF
scheme to the Random scheme. We use the following heuristic approach
to define the phase transition point $\kappa$  at which the
switching takes place.  We define $\kappa\in[1, h]$ to be the
smallest integer that satisfies the following two criteria, where
$h=M-N+1$ is the maximum number of links that can be removed:
\begin{enumerate}
\item $ R_{\kappa}>R_{\kappa+j}$, $j=1,\cdots,l-1$, where $l$ is a window size for the number of
links removed;
\item $R_{\kappa+l-1}<\frac{R_{\kappa}}{2} $.
\end{enumerate}

In Fig. \ref{rc}, we also plot the Hybrid scheme with $l=20$, an
empirical value for various scale-free networks. It can be seen that
applying Hybrid scheme in the BA network, the phase transition
disappears, and $R_c$ decreases gradually after the corresponding
phase transition point in the LBF scheme.

\begin{figure}[htb]
\centering
\includegraphics[width=8cm]{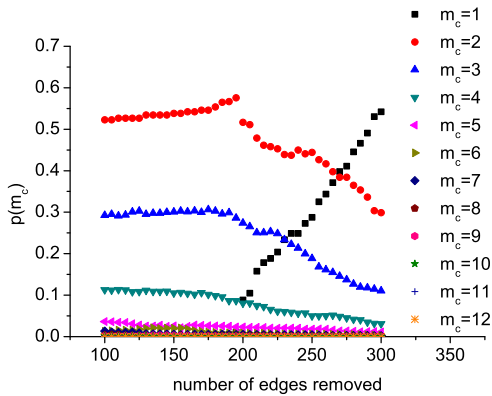}
\caption{The probability distribution of the minimum cut between any
node pairs for a wide range of networks around the phase transition
in the BA network. $m_c$ represents the minimum cut, the $X$-axis
represents the number of links being removed by the LBF scheme, and
the $Y$-axis represents the probability of a node pair whose minimum
cut is $m_c$. \label{mc-distribution}}
\end{figure}

The reason for the occurrence of phase transition in the LBF scheme
can be expressed in the following way. Initially, removing the
critical links (i.e., high edge betweenness links) in scale-free
networks can make the traffic more balanced in the network, avoid
the traffic congestion in the busy links, and consequently enhance
the overall transmission efficiency. However, this effect should
come under the condition that the path diversity is not
significantly affected. When certain amount of these critical links
are removed, continual removal of these critical links will greatly
reduce the path diversity between node pairs. Lacking of path
diversity will make some links unable to be bypassed and congestion
unable to be avoided. To analyze this effect, we use the minimum
cut, $m_c$, between any node pairs as a measure of path diversity,
and plot in Fig. \ref{mc-distribution} the probability distribution
of this measure for a wide range of networks around the phase
transition of the BA network. Note that the minimum cut between two
nodes defines the upper bound of the number of edge disjoint
shortest paths between these two nodes. It can be observed that
before the phase transition, recursive removal of links with the
largest edge betweenness in the BA network has little impact on the
minimum cut between any node pairs. However, this property begin to
change around the phase transition point. Notably, the probability
that a node pair has only a single edge disjoint path between them,
i.e., $m_c$=1, increases very quickly after the phase transition
point. This indicates that around the phase transition, the benefit
of load balancing arising from the removal of critical links begins
to fade away, whereas the negative effect on the network's path
diversity increases abruptly when links are continued to be removed
in this way.

\begin{figure}[htb]
\centering
\includegraphics[width=8cm]{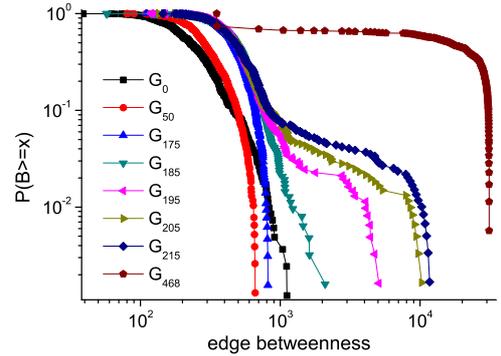}
\caption{The cumulative probability distribution of the edge
betweenness. $G_{i}$ means the network derived by removing $i$ edges
under the LBF scheme. $G_0$ is the original network, $G_{50}$ is the
network with maximum $R_c$, $G_{175}$ is the network before the
phase transition, and $G_{468}$ is the spanning tree.
\label{eb-cdf}}
\end{figure}

Fig. \ref{eb-cdf} also presents the cumulative distribution of edge
betweenness of the original network, the network with maximum $R_c$,
several networks around the phase transition point, and the final
spanning tree of the BA network. It can be seen that just before the
phase transition, the distribution of edge betweenness is relatively
uniform, even more uniform than the original network. This confirms
that at the initial stage of the LBF scheme, removing critical links
can balance the traffic in the network. However, after the phase
transition, the distribution becomes more heavy tailed, reflecting
the performance degradation of the LBF scheme.
%Note that an average of about 400 links in the BA network %(365 links in PFP, 13 in ER)
%have been removed from the original network  before the phase
%transition, whereas only 20 links are removed from the network
%before the phase transition to the network after the phase
%transition. From the original network to the point before phase
%transition, although a significant number of links have been
%removed, the number of node pairs that have only one edge disjoint
%path doesn't show big difference.
%\footnote{Why there are node pairs
%having one edge disjoint path in the original BA network arises from
%the fact that when building the BA network, the nodes in the initial
%network all have degree zero, so some nodes in the generated graph
%may have degree 1. }
%However, though only a small number of links are removed, we see a
%surge of this number before and after the phase transition point.

\begin{figure*}[htb]
\centering
\subfigure[ER]{\includegraphics[width=5.8cm]{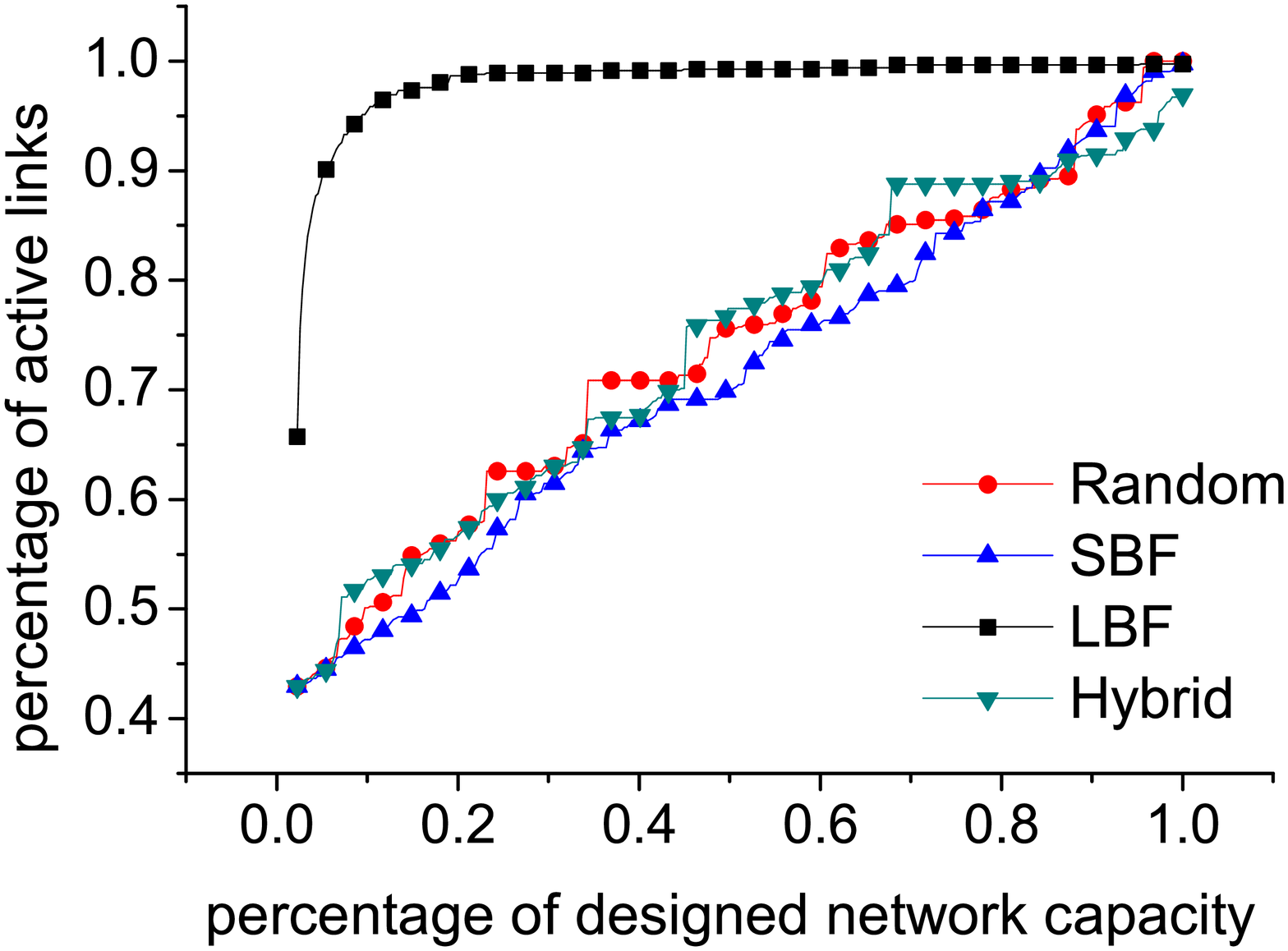}}
\subfigure[BA]{\includegraphics[width=5.8cm]{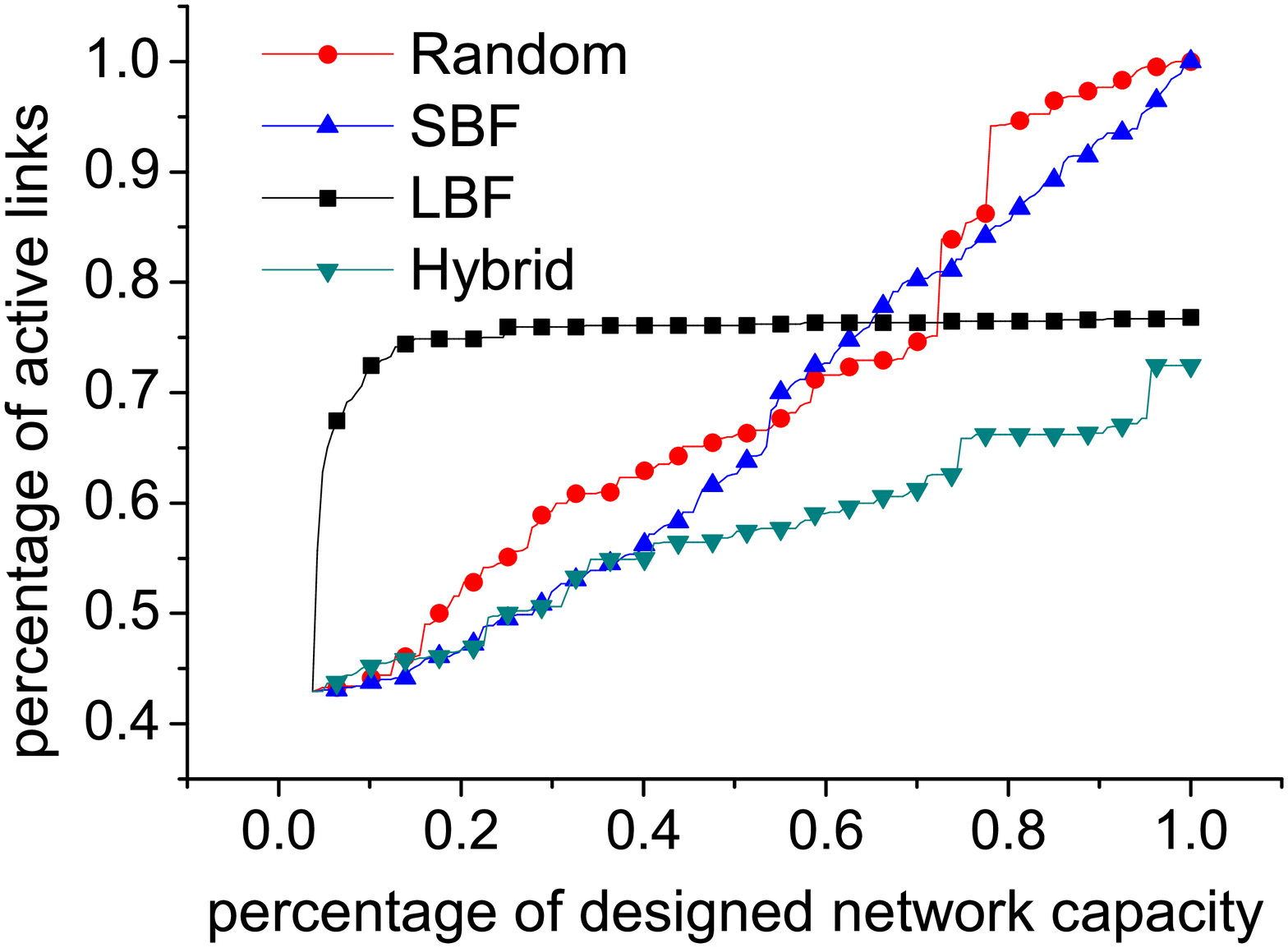}}
%\subfigure[PFP]{\includegraphics[width=8cm]{4c}}
\subfigure[AS3967]{\includegraphics[width=5.8cm]{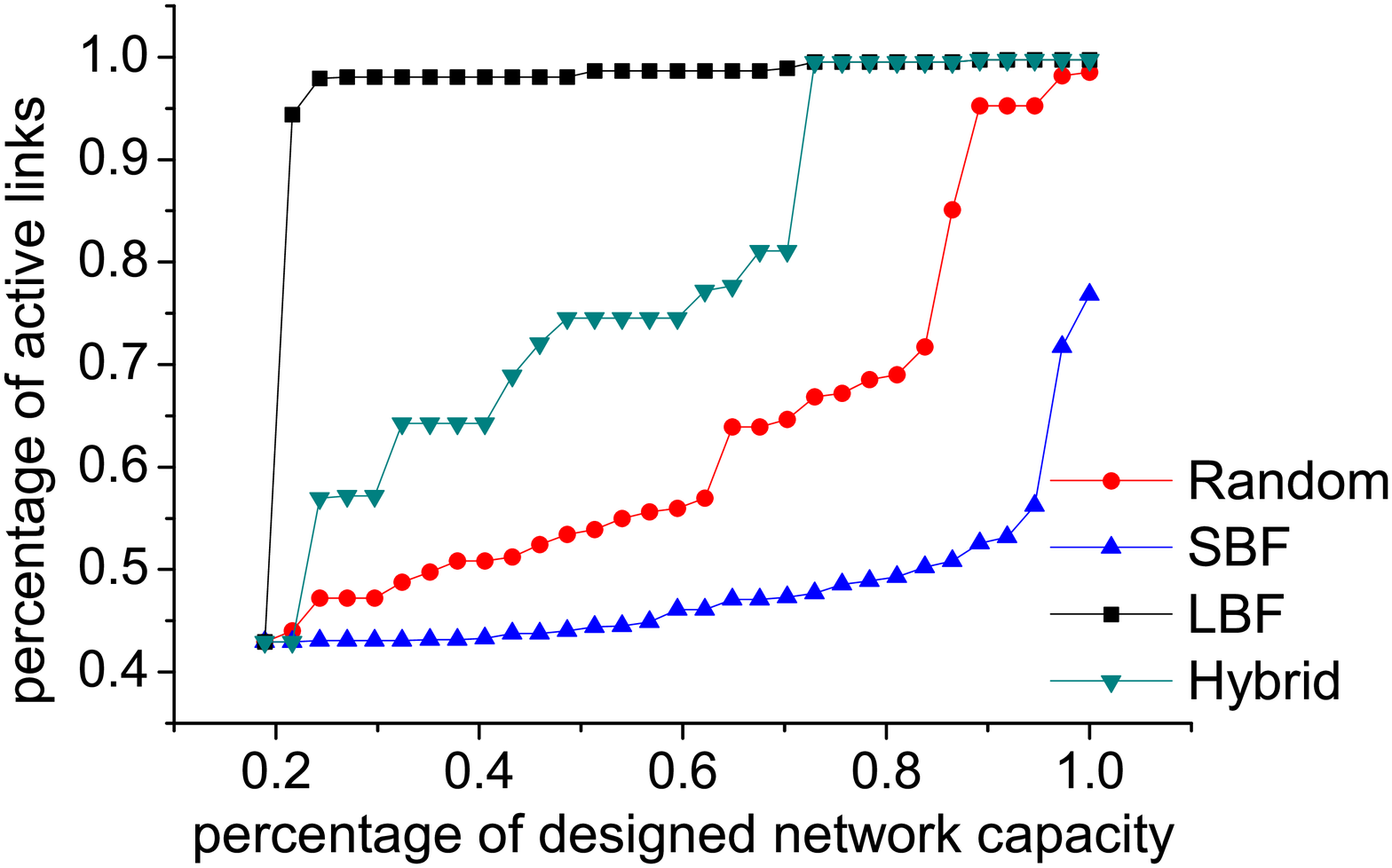}}
\caption{Correlation between the percentage of active links and the
traffic load under different link coordination schemes in different
networks. The $X$-axis is the percentage of traffic load of the
designed network capacity, i.e., $R_c$ of the original
network.\label{correlation} }
\end{figure*}

We also observe clear differences between the real ISP's
router-level Internet topology and the other two synthesized
networks. One difference is that the designed network capacity of
AS3967 is much smaller than other networks, although their network
sizes are the same. This effect is also reported in our previous
work \cite{zhang2010}. Another major difference is that with the SBF
scheme, all the $R_c$ values of the two synthesized networks
decrease steadily, while the $R_c$ value of AS3967 decreases very
slowly until a large portion of the links are removed. On the
contrary, the LBF scheme performs extremely bad. The above
differences indeed all arise from the fact that the router-level
Internet topology is structured in a more hierarchical manner. In
the router-level Internet topology, routers can be coarsely
classified into core routers, edge routers and aggregation routers.
Aggregation routers account for the largest portion of its nodes.
Typically they connect to access routers, which in turn connect to
core routers, forming an apparent hierarchical structure. So,
different aggregation routers that share no common access routers
have to rely on core routers to reach each other. Hence, links
between core routers often have very large edge betweenness,
resulting in relatively small designed network capacity. In the LBF
scheme, these links are the first to be removed. However, removing
these links cannot make the traffic more balanced, but instead, will
put more pressure on the remaining core links, which expresses why
the LBF scheme performs extremely bad. On the other hand,
aggregation routers are often connected to multiple access routers
to improve their robustness for network accessibility and to avoid
single point of failure. These links are provided only for
redundancy or backup purposes, which have very small edge
betweenness values. Consequently, removing these links will not
affect the network's capacity, which expresses why the SBF scheme
shows remarkably good performance in AS3967.

\section{Quantitative Measurement of Energy Savings}
 Assuming the packet injecting rate $R$ decreases from $\lfloor R_0\rfloor$ by
one for each time unit until $\lfloor R_{h}\rfloor$, then given a
link powerdown scheme $\Gamma$, the overall sleeping time units
$SLEEP^{(\Gamma)}$ of these links during the traffic cooling down
period are:
\begin{equation}
SLEEP^{(\Gamma)}=\sum_{R=\lfloor R_h \rfloor}^{\lfloor R_0
\rfloor}{max\{k|R_i\ge R, \forall i\in[1,k]\}}
\end{equation}

The potential energy savings can then be measured by the ratio
between $SLEEP^{(\Gamma)}$ and the maximal active time when all
links are active during the whole period, i.e., $M(\lfloor
R_0\rfloor-\lfloor R_h\rfloor)$, where $M$ is the number of links in
the original network.

Applying the above analysis, Table \ref{table-energy-savings}
presents the average achievable energy savings under different
coordinated link powerdown schemes for the three networks. We
observed that the LBF scheme performs extremely bad on the ER
network and AS3967. During the whole traffic cooling down phase, the
overall achievable energy savings are only less than 2\% and 4\%
respectively. On the ER network, the other three schemes, Random,
SBF and Hybrid, have similar performance. On the BA network, we
found that the Hybrid scheme performs much better than the other
three schemes. While on the AS3967, the SBF scheme outperforms other
schemes significantly.

\begin{table}[h]
\centering
 \caption{Percentage of energy savings that can be
achieved under different link power down schemes on the three
networks. \label{table-energy-savings}}
\begin{tabular}{|c|cccc|}
\hline
Network & Random & SBF & LBF & Hybrid \\
\hline ER &26.0\% &28.6\% &1.7\% &26.1\%\\
BA & 29.9\%& 32.9\% &24.9\% & 43.0\%\\
%PFP & 35.2\% & 39.9\% & 22.3\% & 48.8\% \\
AS3967 & 37.8\% & 53.6\% & 3.2\% &22.0\% \\
 \hline
\end{tabular}
\end{table}

The achievable energy savings critically depend on the correlation
between the number of active links and the traffic load (measured by
the percentage of designed network capacity). Fig. \ref{correlation}
presents the correlation between the two factors. If we treat the
percentage of active links as a function of the traffic load, we
have the following observations:
\begin{itemize}
\item With the Random scheme, the percentage of active links is
an approximately linear function of the traffic load on the two
synthesized networks, whereas a weaker linear correlation is
observed on AS3967.  In all cases, nearly all the links have to be
kept active when the traffic load reaches the designed network
capacity.
\item The SBF scheme is similar to the Random scheme for
the two synthesized networks, however, it shows quite different
performance on AS3967. When applying the SBF scheme on AS3967, only
a small portion of the links have to be kept active for a wide range
of traffic load. Indeed, in this case, the number of active links is
a concave function of the traffic load, which is superior to both
the linear and convex functions in terms of energy savings.
\item With the LBF scheme, a large portion of links should be
kept active for a wide range of traffic load. This exact portion is,
however, dependent on the network topologies. For the ER network and
AS3967, even with very low traffic load, nearly all the links have
to be kept active to satisfy the QoS, while for the BA network, a
constant fraction of the links can be put into sleep for a wide
range of traffic load.
\item The Hybrid scheme performs
similarly to the Random scheme on the ER network, because the phase
transition always takes place at the very beginning of the link
powerdown process. In the BA network, in addition to enable linear
correlation between the number of active links and traffic load, the
Hybrid scheme also requires much fewer links to be kept in active
state when the traffic load reaches the designed network capacity.
\end{itemize}

\section{Conclusion and Discussion}
In this paper, we show that significant energy can be saved if links
are intelligently put into sleep during periods of low traffic load.
Even very simple schemes such as the random link powerdown scheme
can save significant amount of electricity usage. Optimized energy
savings, however, depend not only on the link coordination schemes,
but also on network topologies.

We observe both linear and convex relationships between the number
of active links and the traffic load on the two synthesized
networks, whereas on AS3967, we observe concave relationship between
the above two factors, a notably good relationship for energy
savings. However, AS3967 has much smaller designed network capacity
than the two synthesized networks. From the theoretical point of
view, it is interesting to investigate whether there are network
topologies that simultaneously allow high designed network
capacities and concave relationships between the number of active
links and the traffic load.

As a theoretical work, several practical issues remain to be
addressed in order to realize the potential energy savings. The
first issue is when to power down/up the links in practice? One
possible solution is to leverage the empirical daily traffic profile
to guide the decision. Another important issue is that powering
up/down the links triggers the routing changes. How these schemes
affect the routing table changes and end-to-end delays remains to be
investigated in future practical studies. In practice, those schemes
that enable similar energy savings but trigger the least amount of
routing change events should be favored.

\acknowledgments G. Q. Zhang thanks the Hi-Tech Research and
Development Program of China under Grant No. 2008AA01Z203. G. Q.
Zhang also thanks professor T. Zhou from University of Electronic
Science Technology of China and prefessor S. Zhou from University
College London for their valuable comments on this paper.

\end{document}